\documentstyle[12pt]{article}

\expandafter\ifx\csname amssym.def\endcsname\relax \else\endinput\fi
\expandafter\edef\csname amssym.def\endcsname{%
       \catcode`\noexpand\@=\the\catcode`\@\space}
\catcode`\@=11
\def\undefine#1{\let#1\undefined}
\def\newsymbol#1#2#3#4#5{\let\next@\relax
 \ifnum#2=\@ne\let\next@\msafam@\else
 \ifnum#2=\tw@\let\next@\msbfam@\fi\fi
 \mathchardef#1="#3\next@#4#5}
\def\mathhexbox@#1#2#3{\relax
 \ifmmode\mathpalette{}{\m@th\mathchar"#1#2#3}%
 \else\leavevmode\hbox{$\m@th\mathchar"#1#2#3$}\fi}
\def\hexnumber@#1{\ifcase#1 0\or 1\or 2\or 3\or 4\or 5\or 6\or 7\or 8\or
 9\or A\or B\or C\or D\or E\or F\fi}

\ifcase\@ptsize
  \font\tenmsa=msam10
  \font\sevenmsa=msam7
  \font\fivemsa=msam5
\or
  \font\tenmsa=msam10  scaled \magstephalf
  \font\sevenmsa=msam7 scaled \magstephalf
  \font\fivemsa=msam5  scaled \magstephalf
\or
  \font\tenmsa=msam10  scaled \magstep1
  \font\sevenmsa=msam7 scaled \magstep1
  \font\fivemsa=msam5  scaled \magstep1
\fi

\newfam\msafam
\textfont\msafam=\tenmsa
\scriptfont\msafam=\sevenmsa
\scriptscriptfont\msafam=\fivemsa
\edef\msafam@{\hexnumber@\msafam}
\mathchardef\dabar@"0\msafam@39
\def\dashrightarrow{\mathrel{\dabar@\dabar@\mathchar"0\msafam@4B}}
\def\dashleftarrow{\mathrel{\mathchar"0\msafam@4C\dabar@\dabar@}}

\def\ulcorner{\delimiter"4\msafam@70\msafam@70 }
\def\urcorner{\delimiter"5\msafam@71\msafam@71 }
\def\llcorner{\delimiter"4\msafam@78\msafam@78 }
\def\lrcorner{\delimiter"5\msafam@79\msafam@79 }
\def\yen{{\mathhexbox@\msafam@55 }}
\def\checkmark{{\mathhexbox@\msafam@58 }}
\def\circledR{{\mathhexbox@\msafam@72 }}
\def\maltese{{\mathhexbox@\msafam@7A }}

\ifcase\@ptsize
  \font\tenmsb=msbm10
  \font\sevenmsb=msbm7
  \font\fivemsb=msbm5
\or
  \font\tenmsb=msbm10  scaled \magstephalf
  \font\sevenmsb=msbm7 scaled \magstephalf
  \font\fivemsb=msbm5  scaled \magstephalf
\or
  \font\tenmsb=msbm10  scaled \magstep1
  \font\sevenmsb=msbm7 scaled \magstep1
  \font\fivemsb=msbm5  scaled \magstep1
\fi

\newfam\msbfam
\textfont\msbfam=\tenmsb
\scriptfont\msbfam=\sevenmsb
\scriptscriptfont\msbfam=\fivemsb
\edef\msbfam@{\hexnumber@\msbfam}
\def\Bbb#1{{\fam\msbfam\relax#1}}
\def\widehat#1{\setbox\z@\hbox{$\m@th#1$}%
 \ifdim\wd\z@>\tw@ em\mathaccent"0\msbfam@5B{#1}%
 \else\mathaccent"0362{#1}\fi}
\def\widetilde#1{\setbox\z@\hbox{$\m@th#1$}%
 \ifdim\wd\z@>\tw@ em\mathaccent"0\msbfam@5D{#1}%
 \else\mathaccent"0365{#1}\fi}

\ifcase\@ptsize
  \font\teneufm=eufm10
  \font\seveneufm=eufm7
  \font\fiveeufm=eufm5
\or
  \font\teneufm=eufm10   scaled \magstephalf
  \font\seveneufm=eufm7  scaled \magstephalf
  \font\fiveeufm=eufm5   scaled \magstephalf
\or
  \font\teneufm=eufm10   scaled \magstep1
  \font\seveneufm=eufm7  scaled \magstep1
  \font\fiveeufm=eufm5   scaled \magstep1
\fi

\newfam\eufmfam
\textfont\eufmfam=\teneufm
\scriptfont\eufmfam=\seveneufm
\scriptscriptfont\eufmfam=\fiveeufm

\csname amssym.def\endcsname

\expandafter\ifx\csname pre amssym.tex at\endcsname\relax \else \endinput\fi
\expandafter\chardef\csname pre amssym.tex at\endcsname=\the\catcode`\@
\catcode`\@=11
\newsymbol\boxdot 1200
\newsymbol\boxplus 1201
\newsymbol\boxtimes 1202
\newsymbol\square 1003
\newsymbol\blacksquare 1004
\newsymbol\centerdot 1205
\newsymbol\lozenge 1006
\newsymbol\blacklozenge 1007
\newsymbol\circlearrowright 1308
\newsymbol\circlearrowleft 1309
\undefine\rightleftharpoons
\newsymbol\rightleftharpoons 130A
\newsymbol\leftrightharpoons 130B
\newsymbol\boxminus 120C
\newsymbol\Vdash 130D
\newsymbol\Vvdash 130E
\newsymbol\vDash 130F
\newsymbol\twoheadrightarrow 1310
\newsymbol\twoheadleftarrow 1311
\newsymbol\leftleftarrows 1312
\newsymbol\rightrightarrows 1313
\newsymbol\upuparrows 1314
\newsymbol\downdownarrows 1315
\newsymbol\upharpoonright 1316
 
\newsymbol\downharpoonright 1317
\newsymbol\upharpoonleft 1318
\newsymbol\downharpoonleft 1319
\newsymbol\rightarrowtail 131A
\newsymbol\leftarrowtail 131B
\newsymbol\leftrightarrows 131C
\newsymbol\rightleftarrows 131D
\newsymbol\Lsh 131E
\newsymbol\Rsh 131F
\newsymbol\rightsquigarrow 1320
\newsymbol\leftrightsquigarrow 1321
\newsymbol\looparrowleft 1322
\newsymbol\looparrowright 1323
\newsymbol\circeq 1324
\newsymbol\succsim 1325
\newsymbol\gtrsim 1326
\newsymbol\gtrapprox 1327
\newsymbol\multimap 1328
\newsymbol\therefore 1329
\newsymbol\because 132A
\newsymbol\doteqdot 132B
 
\newsymbol\triangleq 132C
\newsymbol\precsim 132D
\newsymbol\lesssim 132E
\newsymbol\lessapprox 132F
\newsymbol\eqslantless 1330
\newsymbol\eqslantgtr 1331
\newsymbol\curlyeqprec 1332
\newsymbol\curlyeqsucc 1333
\newsymbol\preccurlyeq 1334
\newsymbol\leqq 1335
\newsymbol\leqslant 1336
\newsymbol\lessgtr 1337
\newsymbol\backprime 1038
\newsymbol\risingdotseq 133A
\newsymbol\fallingdotseq 133B
\newsymbol\succcurlyeq 133C
\newsymbol\geqq 133D
\newsymbol\geqslant 133E
\newsymbol\gtrless 133F
\newsymbol\sqsubset 1340
\newsymbol\sqsupset 1341
\newsymbol\vartriangleright 1342
\newsymbol\vartriangleleft 1343
\newsymbol\trianglerighteq 1344
\newsymbol\trianglelefteq 1345
\newsymbol\bigstar 1046
\newsymbol\between 1347
\newsymbol\blacktriangledown 1048
\newsymbol\blacktriangleright 1349
\newsymbol\blacktriangleleft 134A
\newsymbol\vartriangle 134D
\newsymbol\blacktriangle 104E
\newsymbol\triangledown 104F
\newsymbol\eqcirc 1350
\newsymbol\lesseqgtr 1351
\newsymbol\gtreqless 1352
\newsymbol\lesseqqgtr 1353
\newsymbol\gtreqqless 1354
\newsymbol\Rrightarrow 1356
\newsymbol\Lleftarrow 1357
\newsymbol\veebar 1259
\newsymbol\barwedge 125A
\newsymbol\doublebarwedge 125B
\undefine\angle
\newsymbol\angle 105C
\newsymbol\measuredangle 105D
\newsymbol\sphericalangle 105E
\newsymbol\varpropto 135F
\newsymbol\smallsmile 1360
\newsymbol\smallfrown 1361
\newsymbol\Subset 1362
\newsymbol\Supset 1363
\newsymbol\Cup 1264
 
\newsymbol\Cap 1265
 
\newsymbol\curlywedge 1266
\newsymbol\curlyvee 1267
\newsymbol\leftthreetimes 1268
\newsymbol\rightthreetimes 1269
\newsymbol\subseteqq 136A
\newsymbol\supseteqq 136B
\newsymbol\bumpeq 136C
\newsymbol\Bumpeq 136D
\newsymbol\lll 136E
 
\newsymbol\ggg 136F
 
\newsymbol\circledS 1073
\newsymbol\pitchfork 1374
\newsymbol\dotplus 1275
\newsymbol\backsim 1376
\newsymbol\backsimeq 1377
\newsymbol\complement 107B
\newsymbol\intercal 127C
\newsymbol\circledcirc 127D
\newsymbol\circledast 127E
\newsymbol\circleddash 127F
\newsymbol\lvertneqq 2300
\newsymbol\gvertneqq 2301
\newsymbol\nleq 2302
\newsymbol\ngeq 2303
\newsymbol\nless 2304
\newsymbol\ngtr 2305
\newsymbol\nprec 2306
\newsymbol\nsucc 2307
\newsymbol\lneqq 2308
\newsymbol\gneqq 2309
\newsymbol\nleqslant 230A
\newsymbol\ngeqslant 230B
\newsymbol\lneq 230C
\newsymbol\gneq 230D
\newsymbol\npreceq 230E
\newsymbol\nsucceq 230F
\newsymbol\precnsim 2310
\newsymbol\succnsim 2311
\newsymbol\lnsim 2312
\newsymbol\gnsim 2313
\newsymbol\nleqq 2314
\newsymbol\ngeqq 2315
\newsymbol\precneqq 2316
\newsymbol\succneqq 2317
\newsymbol\precnapprox 2318
\newsymbol\succnapprox 2319
\newsymbol\lnapprox 231A
\newsymbol\gnapprox 231B
\newsymbol\nsim 231C
\newsymbol\ncong 231D
\newsymbol\diagup 231E
\newsymbol\diagdown 231F
\newsymbol\varsubsetneq 2320
\newsymbol\varsupsetneq 2321
\newsymbol\nsubseteqq 2322
\newsymbol\nsupseteqq 2323
\newsymbol\subsetneqq 2324
\newsymbol\supsetneqq 2325
\newsymbol\varsubsetneqq 2326
\newsymbol\varsupsetneqq 2327
\newsymbol\subsetneq 2328
\newsymbol\supsetneq 2329
\newsymbol\nsubseteq 232A
\newsymbol\nsupseteq 232B
\newsymbol\nparallel 232C
\newsymbol\nmid 232D
\newsymbol\nshortmid 232E
\newsymbol\nshortparallel 232F
\newsymbol\nvdash 2330
\newsymbol\nVdash 2331
\newsymbol\nvDash 2332
\newsymbol\nVDash 2333
\newsymbol\ntrianglerighteq 2334
\newsymbol\ntrianglelefteq 2335
\newsymbol\ntriangleleft 2336
\newsymbol\ntriangleright 2337
\newsymbol\nleftarrow 2338
\newsymbol\nrightarrow 2339
\newsymbol\nLeftarrow 233A
\newsymbol\nRightarrow 233B
\newsymbol\nLeftrightarrow 233C
\newsymbol\nleftrightarrow 233D
\newsymbol\divideontimes 223E
\newsymbol\varnothing 203F
\newsymbol\nexists 2040
\newsymbol\Finv 2060
\newsymbol\Game 2061
\newsymbol\mho 2066
\newsymbol\eth 2067
\newsymbol\eqsim 2368
\newsymbol\beth 2069
\newsymbol\gimel 206A
\newsymbol\daleth 206B
\newsymbol\lessdot 236C
\newsymbol\gtrdot 236D
\newsymbol\ltimes 226E
\newsymbol\rtimes 226F
\newsymbol\shortmid 2370
\newsymbol\shortparallel 2371
\newsymbol\smallsetminus 2272
\newsymbol\thicksim 2373
\newsymbol\thickapprox 2374
\newsymbol\approxeq 2375
\newsymbol\succapprox 2376
\newsymbol\precapprox 2377
\newsymbol\curvearrowleft 2378
\newsymbol\curvearrowright 2379
\newsymbol\digamma 207A
\newsymbol\varkappa 207B
\newsymbol\Bbbk 207C
\newsymbol\hslash 207D
\undefine\hbar
\newsymbol\hbar 207E
\newsymbol\backepsilon 237F
\catcode`\@=\csname pre amssym.tex at\endcsname
\def\Box{\hbox{\vrule height1ex\kern-0.4pt
\vbox to 1ex{\hrule width1ex\vfil\hrule width1ex}\kern-0.4pt\vrule height1ex}}
\newcommand{\sqr}[2]{{{\vcenter{\vbox{\hrule height.#2pt
\hbox{\vrule width.#2pt height#1pt \kern#1pt
\vrule width.#2pt}
\hrule height.#2pt}}}}}

\newcommand{\be}{\begin{equation}}

\newcommand{\ee}{\end{equation}}
\newcommand{\al}{\alpha}

\newcommand{\lm}{\lambda}

\newcommand{\ta}{\tau}
\newcommand{\ph}{\phi}

\newcommand{\phv}{\varphi}

\newcommand{\ps}{\psi}

\newcommand{\om}{\omega}

\newcommand{\C}{{\Bbb C}}

\renewcommand{\H}{\mbox{$\cal H$}}

\newcommand{\R}{{\Bbb R}}

\newcommand{\notp}{p \kern-.48em /}

\newcommand{\bea}{\begin{eqnarray}}
\newcommand{\eea}{\end{eqnarray}}

\newcommand{\half}{\mbox{\footnotesize $\frac{1}{2}$}}

\topmargin = - 0.5 cm
 \newcommand{\arctanh}{{\rm arctanh}}
\newcommand{\etal}{et.al.\ }
\newcommand{\bi}{\bf}
\newcommand{\Sc}{S_{\rm cov}}
\newcommand{\Hc}{{\cal H}_{\rm cov}}
\newcommand{\bcal}{\cal}
\newcommand{\JMP}{{\it J. Math. Phys.}}
\newcommand{\CQG}{{\it Class. Quant. Grav.}}
\begin{document}
 \setlength{\baselineskip}{1.5\baselineskip}
\thispagestyle{empty}
\title{Against the Wheeler-DeWitt equation}
\author{ N P Landsman\thanks{ E.P.S.R.C. Advanced Research Fellow}\\
  Department of Applied Mathematics and Theoretical Physics\\ University of
Cambridge\\ Silver Street, Cambridge CB3 9EW, U.K. }
\date{\today}
\maketitle
 \begin{abstract}
The ADM approach to canonical general relativity combined with Dirac's
method of quantizing constrained systems leads to the Wheeler-DeWitt
equation.
A number of mathematical as well as physical difficulties that arise
in connection with this equation may be circumvented if one employs a
covariant Hamiltonian method in conjunction with a recently
developed, mathematically rigorous technique to quantize constrained systems
using Rieffel
induction. The classical constraints are cleanly separated into four
components of a covariant momentum map coming from the diffeomorphism
group of spacetime, each of which is linear in the canonical momenta,
plus a single finite-dimensional quadratic constraint that arises in
any theory, parametrized or not.

The new quantization method is carried through in a minisuperspace
example, and is found to produce a ``wavefunction of the universe".
This differs from the proposals of both Vilenkin and  Hartle-Hawking for a
closed
FRW universe, but happens to coincide with the latter in the open case.
\end{abstract}
\newpage
\section{Introduction}
Dirac's theory of constrained systems in classical mechanics (cf.\ Gotay \etal
(1978) for a modern
geometric formulation) consists of two steps. Firstly, the constraints
$\Phi_a=0$ are imposed on
the phase space $S$ of the unconstrained system, singling out the constraint
hypersurface $C\subset
S$ as their solution space. Secondly, one forms the quotient $S^0=C/{\cal F}_0$
of $C$ by the
foliation ${\cal F}_0$ defined by the null directions of the induced symplectic
form $i^*\om$ on $S$
(here $\om$ is the symplectic form on $S$ and $i$ is the injection of $C$ into
$S$). This second
step (which is absent if all constraints are second class) identifies
physically equivalent points
on $C$. If $S=T^*Q$ is  a cotangent bundle and the constraints are components
of a momentum map
$\Phi$ (cf.\ Marsden and Ratiu 1994) derived from an action of a group $G$ on
$Q$, then all
$\Phi_a$ are necessarily {\em linear} in the momenta. The so-called
Marsden-Weinstein quotient
$\Phi^{-1}(0)/G$ then coincides with the reduced phase space.

In canonical (ADM) classical gravity (cf.\ Fischer and Marsden 1979, Isham
1993, Kucha\v{r} 1992) the
configuration space $Q$ is taken to be the space of Riemannian 3-metrics
(subject to certain
regularity conditions) on a (Cauchy) hypersurface $\Sigma$ (here assumed to be
compact) in space-time
$X$. The $\Phi_a$ are the (super) Hamiltonian- and momentum constraints ${\bcal
H}_{\al}^{\bi x}$ (one
for each point ${\bi x}$ of $\Sigma$; $\al=0,1,2,3$), the first of which are
{\em quadratic}  in the
canonical momenta.   Also, the Poisson bracket of two Hamiltonian constraints
is proportional to the
inverse of the three-metric, which makes it impossible to formulate these
constraints as the
pull-back of a Poisson morphism $\Phi$ from $S$ into {\em any} Poisson manifold
$P$. (The
super-momentum constraints, which are linear in the canonical momenta,
 are of such a nature, with $P$ being
the dual of the Lie algebra of ${\rm Diff}(\Sigma)$ and $\Phi$ the equivariant
momentum map
corresponding to the natural action of this group on $S$.) In particular, the
connection between the
Hamiltonian constraints and ${\rm Diff}(X)$ is obscure. Time-evolution is
generated by the
constraints smeared with  arbitrary functions $N({\bi x},t)$; hence the
dynamical evolution takes
place along some of the null directions of $i^*\om$, and collapses to no
evolution whatsoever on the
physical phase space $S^0$. With obvious modifications, this discussion applies
to the Ashtekar
variables as well.

In quantizing constrained systems, the two-step classcial procedure is replaced
by a single step.
Dirac (1964) singled out the {\em first} step: if $\cal H$ is the Hilbert space
of states of the
unconstrained system (that is, the quantization of $S$) and $\hat{\Phi}_a$ are
self-adjoint
operators on $\cal H$ quantizing the classical constraints, then the quantum
analogue $\H_D$ of the
physical phase space $S^0$ is defined as $\H_D=\{\psi\in\H| \hat{\Phi}_a\ps=0\,
\forall a\}$. This
space then inherits the inner product from $\H$, and is a Hilbert space in its
own right. In other
words, one imposes the constraints and that's it. This only makes sense if all
constraints have 0  as
discrete eigenvalue, with common eigenspace. This condition is rarely satisfied
in practice, and this
has led to certain modifications  of the Dirac proposal (Ashtekar and Tate
1994,  H\'{a}ji\v{c}ek
1994, Ashtekar \etal\ 1995, Marolf 1995b),
in which one solves the constraints on a bigger space than $\H$. One thereby
loses the inner product,
and the main problem is then to construct an inner product on the solution
space from scratch.
In some examples, such methods lead to the same result as our approach, but in
the former
one still tries to mimick the {\em first} step of the classical reduction
process.
Moreover, it is easy to devise examples where such rigged Hilbert space
techniques will not lead to
the desired answer, see below.

Applied to canonical gravity (Ehlers and
Friedrich 1994), the Dirac procedure leads to the Wheeler-DeWitt equation, in
which
the quantized Hamiltonian constraints are imposed on the wavefunction $\ps$,
which is a function(al)
of the three-metric on $\Sigma$ (or of the corresponding Ashtekar variables).
Quite apart from the
fact to what extent $\H$ is well-defined  (for considerable progress in this
direction see Ashtekar
\etal\ 1994, 1995), this approach meets formidable obstacles (Kucha\v{r}  1992,
Isham 1993).

The problems with the Wheeler-DeWitt equation can be traced back to (at least)
two sources:
i) the lack of covariance of the ADM approach, which is especially dangerous in
connection with
quantum field theory, where  fields are not defined at sharp times; ii) the
Dirac quantization
method of constrained systems. It seems that one can do better on both
accounts. Firstly, there
exists a covariant Hamiltonian formulation of classical field theory (Kijowksi
1973, Gotay 1991,
Gotay \etal\ 1993), and secondly the author (Landsman 1995) has recently
formulated a new method of
quantizing constrained systems, which avoids many   problems in the Dirac
approach.
\section{Constrained quantization revisited}
To start with the latter, the main idea is to implement a quantized version of
the {\em second} step
of the classical constraint procedure, rather than the first one. The
constraints then do not have to
be implemented,  avoiding the Wheeler-DeWitt equation altogether. This is done
by modifying
the inner product $(\, ,\,)$ on $\H$ (which is positive definite) into a
positive semi-definite
sesquilinear form $(\, ,\,)_0$. The latter has a nontrivial null space ${\cal
N}_0=\{\ps\in\H|
(\ps,\ps)_0=0\}$, in terms of which the state space $\H^0$ of the constrained
system is simply given
by $\H^0=\H/{\cal N}_0$. Crucially, $\H^0$ inherits the form $(\, ,\,)_0$,
which is now positive
definite since its null vectors have been thrown away. Thus one has a bona fide
inner product on
$\H^0$, which can be used to calculate physical amplitudes and probabilities.

To illustrate the method,
consider a compact group $G$, which acts on $\H$ through a unitary
representation $U$. If the
constraints correspond to the generators $T_a$ of (the Lie algebra of) $G$, the
modified inner
product turns out to be given by (Landsman 1995) $(\ps,\phv)_0=\int_G dg
(U(g)\ps,\phv)$. One
recognizes that the right-hand side coincides with $(P_0\ps,P_0\phv)$, where
$P_0$ is the orthogonal
projector onto the subspace $\H_0$ of $\H$ which transforms trivially under
$G$. Hence ${\cal
N}_0=\H_0^{\perp}$, so that $\H^0=\H_0$, which coincides with $\H_D$ of Dirac.
The reason one can get
away with not explicitly implementing the constraints is that states not
satisfying them are of zero
norm in the modified inner product, and therefore quotient to the null vector
in $\H^0$.
Also, the action of a constraint on a vector maps it into a null vector (in
other words, a gauge
transformation changes a vector only by a null vector).

Had $G$ not been compact, the modified form  would be ill-defined on some of
$\H$, but our method is
easily adapted to such instances: instead of all of $\H$ one finds a dense
subspace $L\subset \H$ on
which $(\, ,\,)_0$ {\em is} defined. The only change in the procedure is then
that the quotient
$L/{\cal N}_0$ is not complete under the induced norm, and has to be completed
to form $\H^0$.
In this way, constraints with continuous spectrum can be handled without any
difficulty (Landsman
1995). The general way of finding  the modified inner product is based on the
mathematical analogy
between the momentum map and generalized Marsden-Weinstein reduction in
symplectic geometry,   and
Hilbert $C^*$-modules and Rieffel  induction (Rieffel 1974) in operator algebra
theory (cf.\ Landsman
1995 for details and references).

For an interesting example which highlights the difference between our Rieffel
induction method and
rigged Hilbert space techniques,  take $G$ a noncompact semisimple Lie group,
and $S=T^*G$ with
momentum map $\Phi$ defined by the natural right-action of $G$ on $S$.
The classical reduced space $\Phi^{-1}(0)/G$ is a point.
Quantization with our method proceeds by taking $\H=L^2(G)$ as the quantization
of $S$, which carries
the right-regular representation $U$ of $G$.
Let
 $(\, ,\,)_0$ on $L=C_c(G)$ be defined as in the compact case. The physical
Hilbert space
comes out as $\H^0=\C$, which is obviously the correct answer. However, the
trivial
  representation of $G$ is not weakly contained
in the regular representation on $L^2(G)$, so using the standard rigged Hilbert
space ${\cal S}(G)\subset L^2(G) \subset {\cal S}(G)'$, where ${\cal
S}(G)$ is the Harish-Chandra Schwartz space of $G$,
 would produce an empty physical state space. (Note that the word `rigged' as
used in
Landsman (1995) to describe the modified inner product has nothing to do with
the same word
occurring in `rigged Hilbert space'.)

Let $V\ps\in\H^0$ be the image of $\ps\in L$ in the quotient $L/{\cal N}_0$, so
that
$V:L\rightarrow \H^0$ satisfies $(V\ps,V\phv)_{{\cal H}^0}=(\ps,\phv)_0$.
In our examples, $\H^0$ is of the form $L^2(\R^n)$ (in momentum space), and $V$
has the form
$
(V\ps)({\bi p})=(\ps,f_{\bi p})
$ for a family of functions $f_{\bi p}$ lying in some appropriate dual of $L$;
the r.h.s.\ can be
calculated as if it were the inner product in $\H$, although $f_{\bi p}$ is not
in $\H$, cf.\
Poerschke and Stolz 1993.

 Observables are those self-adjoint operators $A$ on $\H$ which satisfy
$(A\ps,\phv)_0=(\ps,A\phv)_0$ for all $\ps,\phv$ in $L$; this condition is the
quantum version of the
classical condition that $\{f,\Phi_a\}$ vanish on $C$. In either case, the
point is that observables
are well-defined on the reduced spaces $\H^0$ or $S^0$, respectively. The
physical action $\pi^0$
on $\H^0$ is given by $\pi^0(A)V\ps=VA\ps$.

\section{Covariant Hamiltonian method}
For our quantization method to apply, the
classical reduced space has to have the form of a generalized Marsden-Weinstein
quotient.
This renders
our method inapplicable to ADM gravity. However, the covariant Hamiltonian
approach advertised above
creates a context in which our technique does apply. For the parametrized
particle and minisuperspace
examples we will have to deal with noncompact abelian groups, for which the
Dirac method breaks down.
Yet our method works, the modified inner product  being given by an expression
similar to that in the
compact group example above, with due care taken about the choice of the dense
subspace $L$.

To see how the covariant method operates in a system without constraints,
consider an unparametrized
relativistic particle, with Lagrangian $L=-m\sqrt{1-\dot{\bi x}^2}$, where
${\bi x}=(x^1,x^2,x^3)$
and $\dot{\bi x}=d{\bi x}/dt$. The covariant phase space $\Sc$ of this model
(Kijowski 1973) is ${\Bbb
R}^8$, with co-ordinates $\{t,p,x^i,p^i\}$. The covariant Legendre transform
(Gotay 1991, Gotay
\etal\ 1993) leads to the primary constraint $\Phi=p-H=0$, where $H=\sqrt{{\bi
p}^2+m^2}\equiv\om_{\bi p}$. (Here $H$ happens to coincide with the usual
Hamiltonian, but in field
theory it would be a covariant generalization thereof.) The symplectic form on
$\Sc$ is
$\om=-dp\wedge dt +dp^i\wedge dx^i$. The solutions to the equations of motion
are precisely tangent
to the null direction of $i^*\om$ at any given point. The reduced phase space
$\Sc^0$ of $\Sc$ with
respect to the constraint $p-H$ is the usual phase space $T^*{\Bbb R}^3$. The
co-ordinates
$p$ and $t$ have been eliminated by the reduction procedure. Time evolution on
$S^0$ may be described
by any observable acting as a Hamiltonian on $\Sc^0$, such as $H$.

To quantize, we replace $\Sc$ by the covariant Hilbert space $\Hc=L^2(\R^4)$
(with $\ps$ a function
of $(t,{\bi x})$). According to our method, we have
$$(\ps,\phv)_0=\int_{-\infty}^{\infty} d\lm
(\exp(-i\lm\hat{\Phi})\ps,\phv).$$  We find
$\H^0=L^2(\R^3; d^3{\bi p}/(2\pi)^3)$, and $V$ defined by $f_{\bi p}(t,{\bi
x})=\exp(-i\om_{\bi
p}t+i{\bi p}\cdot{\bi x})$. These functions are a complete set of generalized
solutions of the
constraints, which explains in part why alternative methods (e.g.,
H\'{a}ji\v{c}ek 1994) are
equally well capable of handling this example. This is not generic, however
(cf.\ the last example
below).  \section{The Polyakov particle}
To mimick minisuperspace, we now treat the relativistic particle in the
parametrized  Polyakov
formalism (cf.\ Gotay \etal 1993 for the classical part).
The Lagrangian is
$$L=-\half\sqrt{g}(g^{00}\dot{x}^{\mu} \dot{x}_{\mu}+m^2),$$
 where
$\dot{x}=dx/d\tau$, and $g=g_{00}$ is a metric on the real line. The covariant
phase space is
parametrized by $\{\ta,p,x^{\mu},p_{\mu},g,\pi\}$, subject to the primary
constraints
$p-H=0$, with $$H=-\half\sqrt{g}(p_{\mu}p^{\mu}-m^2),$$ and $\pi=0$.
The symplectic form on $\Sc$ is $\om=-dp\wedge dt-dp_{\mu}\wedge
dx^{\mu}+d\pi\wedge dg$.
According to the general method of Gotay \etal\ (1993) to handle parametrized
theories, the secondary
constraints are picked up  by the covariant momentum map, in this case with
respect to the action of
${\rm Diff}^+(\R)$, under which the Lagrangian is covariant. If $N(\ta)d/d\tau$
is an arbitary element
of the Lie algebra of this group (that is, a vector field on $\R$), this
momentum map is given by
$\Phi_N=-pN-2\pi g \dot{N}$, which is linear in the canonical momenta, as it
should, and generates
diffeomorphisms in the appropriate manner.

The complete set of constraints, then, is $p-H=\pi=\Phi_N=0$ (the latter for
all $N$). In full
general relativity, there would be analogous constraints, with $g$ and $\pi$
replaced by the lapse
and shift components of the metric, and the diffeomorphism constraints now
being of the form
$\int_{\Sigma} \sigma^*\Phi_N=0$ (here $\sigma$ is a section of the covariant
primary constraint
bundle (Gotay \etal\ 1993), and $\Sigma$ is a Cauchy surface). In any case, the
diffeomorphism
constraints are linear in the canonical momenta also in that case.
The primary constraints, i.e., $\pi=0$ and $p=H$, are imposed on the {\em
finite-dimensional} covariant phase space. It is only when $p=H$ is substituted
into the
diffeomorphism constraints (and evaluated at a fixed time) that one quarter  of
the latter become
quadratic in the canonical momenta, ruining their key property of generating
four-dimensional
diffeomorphisms, as well as the property that their Poisson algebra represents
the Lie algebra of the
group. Such a substitution would, in fact, lead to the ADM formalism.

In the covariant Hamiltonian formalism, on the other hand, the constraints are
of the type we can
handle with our quantization method, since the diffeomorphism constraints are
essentially the
(covariant) momentum map, and the finite-dimensional constraint $p=H$, which is
typical to the
covariant method, and independent of the diffeomorphism invariance,  can be
thought of as a momentum
map for an action of $\R$. Also, the formalism clearly distinguishes between
the lapse $g$ and the
infinitesimal diffeomorphism
 $N(\ta)d/d\tau$.

In the present case, instead of reducing with respect to  ${\rm Diff}^+({\R})$,
the same reduced phase
space is obtained if we reduce with respect to an arbitrary one-parameter
subgroup generated by
some $N$, as long as $N>0$.  Also, since $g>0$, we can rewrite the constraints
as $\Phi_1=p=0$,
$\Phi_2=\pi=0$, and $\Phi_3=-\half(p_{\mu}p^{\mu}-m^2)=0$.
Having reduced by $\Phi_1$ and $\Phi_2$,  we then simply have to reduce
$T^*\R^4$ by $\Phi_3$, which of course leads to the correct
result for the reduced phase space; the restriction to $p_0>0$ has to be
imposed by hand (cf.\
Landsman 1995).

The quantization of $\Sc$ is $\Hc=L^2(\R^6)$, with $\ps$ a function of
$(\ta,x^{\mu},\log(g))$.
The modified inner product is given by
$$ (\ps,\phv)_0=\int d\lm_1d\lm_2d\lm_3
(\exp[-i\sum_{a=1}^3\lm_a\hat{\Phi}_a]\ps,\phv),$$
defined on $\ps,\phv\in L=C_c(\R^6)$.
 Imposing $p_0>0$, it follows that the physical Hilbert space $\H^0$ and the
map $V$ are
as in the previous treatment
 (the  $f_{\bi p}$ are now regarded as   functions of
$(\ta,x^{\mu},\log(g))$, which happen to be independent of $\ta$ and $g$).
Again, one easily verifies
that observables such as the generators of the Poincar\'{e} group act in the
correct way.
\section{A minisuperspace example}
We finally turn to a simple minisuperspace example (cf.\ Halliwell 1990). The
Lagrangian is $$
L=-\half\sqrt{g}e^{3\al}(g^{00}(\dot{\al}^2-\dot{\phi}^2)-\kappa e^{-2\al}), $$
which describes an FRW universe with radius $\exp(\al)$, filled with a
homogeneous massless scalar
field $\phi$ (here $\kappa=0,-1$ (open), or $+1$ (closed)). The covariant phase
space $\Sc$ is
parametrized by  $\{\ta,p,\al,p_{\al},\phi,p_{\phi},g,\pi\}$. The
constraints are   as in the previous example, except  that the Hamiltonian is
now given by $$
H=\half\sqrt{g}e^{-3\al}(-p_{\al}^2+p_{\phi}^2-\kappa e^{4\al}). $$ After
elimination of $\ta$ and
$g$, the physical phase space is obtained by reducing $T^*\R^2$ by the
constraint $H=0$.

In a suitable
parametrization (of which the reduced space is evidently independent), which
corresponds to putting
$\Phi_3= \half (-p_{\al}^2+p_{\phi}^2-\kappa \exp(4\al))$, for $\kappa= 1$ the
flow generated by
this constraint is given by $\phi(\lm)=\phi(0)+p_{\phi}\lm$,
$p_{\phi}(\lm)=p_{\phi}(0)$;
$\al(\lm)=\half\log[\sqrt{2E}/\cosh(2\sqrt{2E}(\lm-\lm_0))]$,
$p_{\al}(\lm)=\sqrt{2E}\tanh(2\sqrt{2E}(\lm-\lm_0))$, where
$E=\half(p_{\al}(0)^2+\kappa
\exp(4\al(0)))$ and $\lm_0$ is determined by $\al(0)=\half
\log[\sqrt{2E}/\cosh(2\sqrt{2E}\lm_0)]$
(or $p_{\al}(\lm_0)=0$). Observables are, for example, the functions
$F_1=p_{\phi}$ and \\
$F_2=\phi-p_{\phi}\, (\arctanh(p_{\al}/\sqrt{2E}))/2\sqrt{2E}$, which in fact
project to canonical
co-ordinates $f_1,f_2$ on the reduced space $S^0$.  These are observables of
DeWitt type, e.g.,
$F_2(z)$ is the value of $\phi(z(\lm))$ at the $\lm$ for which
$p_{\al}(z(\lm))=0$, $z\in T^*\R^2$. For $\kappa =-1$ the above
expression for the flow is formally correct as well, but in that case
the classical motion is incomplete at infinity and a special
interpretation is necessary (cf.\ Feinberg and Peleg 1995 for similar
cases).

The idea of taking the matter field $\phi$  (which is not an observable) as a
time  variable (cf.\ Isham 1993) is implemented in a parametrization-invariant
way by using $f_1$ as
the physical Hamiltonian of the model. If $a_0(z)$ is the value of
$\exp(\al(z(\lm)))$ at the $\lm$
for which $\ph(z(\lm))=0$, then a physical prediction of the model would be the
time evolution
$a_0^2(f_1,f_2,t)=f_1/\cosh(2(f_2+t))$. This describes a recollapsing universe,
cf.\ Marolf 1995a.
\section{The wavefunction of the universe}
We quantize on $\Hc=L^2(\R^4)$, with $\ps$ a function of
$(\ta,\al,\phi,\log(g))$.   The modified
inner product is defined as in the previous example.
For $\kappa =0,1$,  $\hat{\Phi}_3$ is the unique self-adjoint extension of the
usual Schr\"{o}dinger
quantization of $\Phi_3$ on the core $C_c^{\infty}(\R^4)$.  For $\kappa =0$ the
analysis is trivial,
leading to $\H_0^0=\oplus_{\pm}L^2(\R; dk/2\pi)$, and $V^{(0)}=V_+^{(0)}\oplus
V^{(0)}_-$ defined
through $f_{k,\pm}^{(0)}(\ta,\al,\phi,\log(g))=\exp(ik(\al\pm\phi))$. For
$\kappa=1$ the physical
Hilbert space is $\H^0_{ 1}=L^2(\R; dk/2\pi)$, and some functional analysis
(for which cf.\ Leis
1979, Picard 1989)
 shows that one has
$$
f_k^{(1)}(\ta,\al,\phi,\log(g))=\pi^{-1}e^{i\phi k}\sqrt{\sinh(\pi
|k|/2)}K_{i|k|/2}(\half e^{2\al}).
$$
For $\kappa =-1$ the operator $\Phi_3$ is not essentially self-adjoint
on
 $C_c^{\infty}(\R^4)$; its deficiency indices are $(1,1)$, and one
obtains
 a one-parameter family of self-adjoint extensions by specifying
boundary conditions at infinity. For all of these, $\H^0_{- 1}=\H^0_{
1}$,
and the
wavefunction of the universe (see below) turns out not to be
 affected by the particular choice
of the extension. For simplicity  we choose the
boundary condition specified in Picard (1989).
This leads to
$$
f_k^{(-1)}(\ta,\al,\phi,\log(g))=\half e^{i\phi k}\sqrt{{\rm cosech}(\pi
|k|/2)}
(J_{i|k|/2}+J_{-i|k|/2})(\half e^{2\al}).
$$
(Different boundary conditions at infinity would have led to a relative
phase between $J_{i|k|/2}$ and $J_{-i|k|/2}$.)

These are solutions to the Wheeler-DeWitt equations of the model (which do not
lie in $\Hc$), but the
difficulty lies in the fact that now there are other linearly independent
solutions as well. Our
technique gives a  procedure of choosing between them. In alternative
quantization methods,
the boundary conditions (at zero radius) singling out the relevant solution are
lost.

One may compare these solutions with proposed `wavefunctions of the universe'
$\Psi$ (cf.\ Halliwell
1990).
 This comparison is feasible if one has a rationale for letting $k\rightarrow
0$ in
$f_k^{(\kappa)}$, which might be provided if the physical Hamiltonian is chosen
to be
$\hat{p}_{\phi}$, as in the classical case. Up to a constant normalization
factor, the Hartle-Hawking
proposal here gives $I_0$ rather than our $K_0$ for $\kappa =1$, and agrees
with our $J_0$ for
$\kappa =-1$ (see Zhuk 1992). Vilenkin's wavefunction, which is always complex,
differs from ours
as well as Hartle-Hawking's in both cases. But note that according to Halliwell
(1990)   both
proposals are somewhat ambiguous and not always well-defined - in fact, one can
find contradictory
statements in the literature concerning these proposals. Our expansion
functions $f^{(\kappa)}$, on
the other hand, only depend on the domain chosen for the quantum
constraints, which at least for $\kappa=0,1$ is uniquely given (within
the bounds of reason).

\section{References}
\begin{description}
\item[] Ashtekar A, Marolf D and Mour\~{a}o J 1994 {\em Proc.\ C. Lanczos
Centenary Conf.}
ed J D Brown \etal\ (Philadelphia: SIAM) gr-qc/9403042
\item[] Ashtekar A, Lewandowski J, Marolf D,  Mour\~{a}o J and Thiemann T 1995
\JMP\ (in press)
gr-qc/9504018
 \item[]
Ashtekar A and Tate R S 1994   \JMP\ {\bf 35} 6434
\item[] Dirac  P A M 1964 {\it Lectures on
Quantum Mechanics} (New York: Yeshiva University Press)
\item[] Ehlers J and Friedrich H (eds) 1994
{\it Canonical Gravity: From Classical to Quantum, Lecture Notes in Physics}
{\bf 434} (Berlin:
Springer)
\item[] Feinberg J and Peleg Y (1995) {\em Phys. Rev.} {\bf 52} 1988
\item[] Fischer A E and Marsden J E 1979 {\it
General Relativity - An Einstein Centenary Survey} eds S W Hawking and W Israel
(Cambridge: Cambridge
University Press)  p 138-211
\item[] Gotay M J 1991 {\em Mechanics, Analysis and Geometry: 200 Years after
Largrange} ed
 M Francaviglia (Amsterdam: Elsevier)
\item[] Gotay M J, Nester J M and Hinds G 1978 \JMP\ {\bf 19} 2388
\item[]  Gotay M J, Isenberg J, Marsden J E, Montgomery R, with the
collaboration of \'{S}niatycki J
and Yasskin P B 1993 {\em Momentum Maps and Classical Relativistic Fields}, to
appear
\item H\'{a}ji\v{c}ek P 1994
{\it Canonical Gravity: From Classical to Quantum, Lecture Notes in Physics}
{\bf 434}
eds J Ehlers  and H Friedrich (Berlin:
Springer) p 113-149
\item[] Halliwell J J 1990 {\em Proc.\ Jerusalem Winter School on Quantum
Cosmology and Baby
Universes} ed T Piran (Singapore: World Scientific)
 \item[] Isham C
J 1993 {\it Integrable Systems, Quantum Groups and Qauntum Field Theories} ed L
A Ibort and M A
Rodr\'{\i}guez (Dordrecht: Kluwer) p 157-287
\item[] Kijowski J 1973 {\em Commun. Math. Phys.} {\bf
30} 99
 \item[] Kucha\v{r} K V 1992 {\it Proc.\ 4th Canadian Conf.\ on General
Relativity and
Relativistic Astrophysics} ed G Kunstatter \etal (Singapore: World Scientific)
\item[] Landsman N P 1995 {\em J. Geom. Phys.} {\bf 15} 285
\item[] Leis R 1979 {\it Math. Meth. Appl. Sci.} {\bf 1} 114
\item[] Marolf D 1995a  \CQG\ {\bf 12} 1199
\item[] Marolf D 1995b {\em Banach Center Publications} (in press)
gr-qc/9508015
\item[] Marsden J E and Ratiu T S 1994 {\em Introduction to Mechanics and
Symmetry} (Berlin:
Springer)
\item[] Picard R 1989 {\it Hilbert Space Approach to some Classical Transforms}
(Harlow:
Longman)
\item[] Poerschke T and Stolz G 1993 {\it Math. Z.} {\bf 212} 337
\item[] Rieffel M A 1974 {\it Adv. Math.} {\bf 13} 176
\item[] Zhuk A 1992 \CQG\ {\bf 9} 2029
\end{description}
\end{document}